\documentclass{article}

\usepackage{multirow,multicol,amsmath,amssymb,latexsym,float,amsthm,rotating}
\usepackage[numbers,sort&compress]{natbib}
\newcommand{\Nbtab}{M}
\newcommand{\nbtab}{m}
\newcommand{\Nbind}{n}
\newcommand{\nbind}{i}
\newcommand{\Nbvar}{p}
\newcommand{\nbvar}{j}
\newcommand{\rkX}{\min(n -1,\,p)}
\newcommand{\Nbsim}{K}
\newcommand{\nbsim}{k}
\newcommand{\Nbdim}{S}
\newcommand{\nbdim}{s}
\newcommand{\bfX}{\textbf{X}}
\newcommand{\bfE}{\textbf{E}}
\newcommand{\bfV}{\textbf{V}}
\newcommand{\bfU}{\textbf{U}}
\newcommand{\bfW}{\textbf{W}}
\newcommand{\bfM}{\textbf{M}}
\usepackage{geometry}
\def\keywords{\vspace{.5em}
{\textit{Keywords}:\,\relax%
}}

\begin{document}


\begin{center}
{\Large{Multiple imputation for continuous variables using a Bayesian principal component analysis}}
\end{center}
 \vglue-2cm
\begin{center}
  \scshape Vincent \textsc{Audigier}\footnote{Principal corresponding author}, Fran\c cois
  \textsc{Husson}\footnote{Corresponding author} and Julie \textsc{Josse}\footnotemark[2]
\renewcommand{\thefootnote}{\arabic{footnote}}\setcounter{footnote}{0}
\end{center}
\vglue0.3cm
\hglue0.02\linewidth\begin{minipage}{0.9\linewidth}
\begin{center}
Applied Mathematics Department,
Agrocampus Ouest, 65 rue de Saint-Brieuc,
F-35042 RENNES Cedex, France \\
\parbox[t]{0.45\linewidth}{\texttt{audigier@agrocampus-ouest.fr}
\texttt{husson@agrocampus-ouest.fr} \texttt{josse@agrocampus-ouest.fr}}
\end{center}
\end{minipage}

\begin{abstract}
We propose a multiple imputation method  based on principal component analysis (PCA) to deal with incomplete continuous data. To reflect the uncertainty of the parameters from one imputation to the next, we use a Bayesian treatment of the PCA model. Using a simulation study and real data sets, the method is compared to two classical approaches: multiple imputation based on joint modelling and on fully conditional modelling. Contrary to the others, the proposed method can be easily used on data sets where the number of individuals is less than the number of variables and when the variables are highly correlated. In addition, it provides unbiased point estimates of quantities of interest, such as an expectation, a regression coefficient or a correlation coefficient, with a smaller mean squared error. Furthermore, the widths of the confidence intervals built for the quantities of interest are often smaller whilst ensuring a valid coverage.
\end{abstract}


\section{Introduction}
Data with continuous variables are ubiquitous in many fields. For instance in biology, samples are described by the expression of the genes, in chemometrics, components can be described by physico-chemical measurements, in ecology, plants are characterized by traits, etc. Whatever the field, missing values occur frequently and are a key problem in statistical practice. Indeed most statistical methods cannot be applied directly on an
incomplete data set. To deal with this issue, one of the common approaches is to perform single imputation. This consists in imputing missing values by plausible values. It leads to a complete data set that can be analysed by any standard statistical method.

However, single imputation is limited because it does not take into account the uncertainty associated with the prediction of missing values based on observed values. Thus, if we apply a statistical method on the completed data table, the variability of the estimators will be underestimated. To avoid this problem, a first solution is to adapt the procedure to be applied on an incomplete data set. To do this, an Expectation-Maximization (EM) algorithm \citep{Dempster77} combined, for instance, with a Supplemented Expectation-Maximization algorithm \citep{Meng91} could be used to get the maximum likelihood estimates as well as their variance from incomplete data. Note that, the maximum likelihood estimate using these algorithms obviates the necessity for imputation. However it is not always easy to establish these algorithms. 
Another solution is to perform multiple imputation \citep{Rubin87,Little02} which consists in predicting different values for each missing value, which leads to several imputed data sets. The variability across the imputations reflects the variance of the prediction of each missing entry. Then, multiple imputation consists in performing the statistical analysis on each completed data set. Finally, the results are combined using Rubin's rules \citep{Rubin87} to obtain an estimate of parameters and an estimate of their variability taking into account uncertainty due to missing data.

Therefore, a multiple imputation method is based on a single imputation method.
Denoting $\theta$  the parameters of the imputation model, a multiple imputation method requires generating a set of $\Nbtab$ parameters $(\widehat{\theta}_1,\dots, \widehat{\theta}_M$) to reflect the uncertainty in the estimate of the model's parameters. Multiple imputation methods are distinguished in the way the uncertainty is spread using either a bootstrap or a Bayesian approach. The bootstrap approach consists in producing $\Nbtab$ new incomplete data sets and estimating $\theta$ on each bootstrap replication. The Bayesian approach consists of determining a posterior distribution for the model's parameters using a prior distribution and the observed entries. Then the set of parameters $(\widehat{\theta}_1,\dots, \widehat{\theta}_M$) is drawn from the posterior distribution.
There are also two classical ways of performing multiple imputation. The first one is to use an explicit joint model to all variables \citep{Schafer97}. A normal distribution is often assumed on variables which may seem restrictive but is known to be fairly robust with respect to the assumption of normality \citep[p.211-218]{Schafer97}. 
The second way to perform multiple imputation is to use chained equations \citep{vanBuuren06}: a model is defined for each variable with missing data and variables are successively imputed using these models. Typically, imputation is done using the regression model or by predictive mean matching. The chained equations approach is more flexible than the joint modelling, however it requires specifying a model for each variable with missing values, which is quite tedious with a lot of incomplete variables. In addition it may not converge to a stationary distribution if the separate models are not compatible \citep{Besag74}, that is to say that there is no joint distribution for variables with the conditional distributions chosen. More generally, the theoretical properties of chained equations are not well understood and they are a current topic of research \citep{liu14}. Both the joint and conditional methods have their own advantages and drawbacks as investigated recently in \citep{Kropko13}. However, both approaches share the drawback that regression models are rapidly ineffective for data sets where the number of individuals is too low compared to the number of variables or when the variables are highly correlated. Even if some solutions using regularization are available to handle such situations, it is not straightforward to deal with such cases.

Recently, \citep{JosseHusson12} proposed a method of single imputation based on a PCA model. This method gives good results in terms of quality of the imputation when there are linear relationships between variables and also has the advantage of being able to be performed on a data set where the number of individuals is smaller than the number of variables. 
 
We propose to extend it to multiple imputation and we spread the uncertainty of parameters of the PCA imputation model using a Bayesian approach. In Section \ref{Method}, we describe the procedure called BayesMIPCA for multiple imputation based on a Bayesian treatment of the PCA model. Then, in Section \ref{SimulationNorm}, we present a simulation study in which we compare this method to other multiple imputation methods and demonstrate that multiple imputation by the BayesMIPCA method produces little bias and valid confidence intervals under a variety of conditions. Finally, we apply the methods on real data sets.

\section{Method\label{Method}}
\subsection{PCA model}
PCA can be expressed using a fixed effect model \citep{Caussinus86} where the data matrix $\bfX_{\Nbind\times \Nbvar}$ can be decomposed as a signal, denoted $\tilde{\bfX}_{\Nbind\times \Nbvar}$, of low rank $\Nbdim$  considered as known, plus noise denoted $\bfE_{\Nbind\times \Nbvar}$:
\begin{eqnarray}\label{pcamodel}
\bfX_{\Nbind \times \Nbvar}=\tilde{\bfX}_{\Nbind\times \Nbvar}+\bfE_{\Nbind\times \Nbvar}
\end{eqnarray} where
$\bfE=\left(\varepsilon_{\nbind \nbvar}\right)_{1\leq\nbind\leq\Nbind, 1\leq\nbvar\leq\Nbvar}$ with $\varepsilon_{\nbind \nbvar}\sim \mathcal{N}(0,\sigma^2)$. 
The parameters of this model are the elements of $\tilde{\bfX}$ and $\sigma$. 

Imputation under the PCA model requires estimating these parameters from the incomplete data set. 
The method which achieves this is closely related to the one applied on a complete data set.

\subsubsection{PCA on complete data}
PCA consists in finding the matrix $\hat{\bfX}$ with rank $\Nbdim$ which minimizes the least squares criterion $\parallel\hat{\bfX}-\bfX\parallel^2$ with $||\cdot||$ the Frobenius norm. Therefore, $\hat{\bfX}$ corresponds to the least squares estimator of $\tilde{\bfX}$. The solution is obtained using the singular value decomposition (SVD) of the matrix $\bfX$:  $\widehat{\bfX}=\bfU\boldsymbol{\Lambda}\bfV^{\top}$
 where columns of $\bfU_{\Nbind \times \Nbdim}$ are the left singular vectors, $\boldsymbol{\Lambda}_{\Nbdim\times \Nbdim}=diag(\lambda_1,\ldots,\lambda_{\Nbdim})$ is the matrix of the singular values of $\bfX$ and columns of $\bfV_{\Nbvar \times \Nbdim}$  are the right singular vectors. The principal components are given by $\bfU\bf{\Lambda}$ and the loadings are given by $\textbf{V}$. This solution also corresponds to the maximum likelihood estimate of model \eqref{pcamodel}. The expression of the general term of $\hat{\bfX}$ is given by
\begin{eqnarray}\label{reconst_partial}
\hat{x}_{\nbind \nbvar}=\sum_{\nbdim =1}^{\Nbdim}{\sqrt{\lambda_{\nbdim}}u_{\nbind \nbdim}v_{\nbvar \nbdim}}.
\end{eqnarray}
Then $\sigma^2$ is estimated by
\begin{eqnarray}\label{estim_sigma}
{\Hat{\sigma}}^2=\frac{\sum_{\nbind \nbvar}{(x_{\nbind \nbvar}-\hat{x}_{\nbind \nbvar})^2}}{\Nbind \Nbvar-(\Nbvar+\Nbdim(\Nbind-1+\Nbvar-\Nbdim))}
\end{eqnarray}
which corresponds to dividing the sum of the squared residuals by the number of entries minus the number of independent model parameters \citep{Candes09}.

The classical PCA estimator \eqref{reconst_partial}, while providing the best low rank approximation of the data matrix, does not ensure the best recovery of the underlying signal $\tilde{\bfX}$.
Thus, other estimators, obtained from regularized versions of PCA, have been suggested in the literature \citep{shabalin13,Verbanck13,Josse15}.
The rationale is exactly the same as in ordinary regression analysis where the maximum likelihood estimates are not necessarily the best ones in terms of mean squared error (MSE), whereas regularized estimators, although more biased have less variability, which lead to a smaller MSE. 
By redefining the problem as finding the best approximation of the unknown signal $\tilde{\bfX}$ in terms of MSE, instead of finding the best low rank approximation of the data matrix $\bfX$, \cite{Verbanck13} suggested a ridge version of the PCA estimator. We focus on this estimator, since, as we will see later in Section \ref{bayesian_PCA}, it has a straightforward Bayesian
interpretation. This better estimator of $\tilde{\bfX}$ in the sense of the mean squared error criterion is defined as follows. Denoting \begin{eqnarray*}
\hat{x}_{\nbind \nbvar}^{(\nbdim)}={\sqrt{\lambda_{\nbdim}}u_{\nbind \nbdim}v_{\nbvar \nbdim}}
\end{eqnarray*}
the $\nbdim^{th}$ term of the sum (\ref{reconst_partial}), this better estimator is determined by searching  $\left(\phi_s\right)_{1\leq \nbdim \leq \Nbdim}$ in order to minimize
\begin{eqnarray*}
\mathbb{E}\left[\sum_{\nbind,\nbvar}\left(\left(\sum_{\nbdim=1}^{\Nbdim}{\phi_s \Hat{x}_{\nbind \nbvar}^{(\nbdim)}}\right)-\tilde{x}_{\nbind \nbvar}\right)^2\right].
\end{eqnarray*}
Note that a parallel with regression analysis and ridge regression can be drawn. \cite{Verbanck13} showed that $\phi_s$ is given by
\begin{align*}
\phi_s=\frac{\sum_{\nbind,\nbvar}\mathbb{E}\left[\hat x_{\nbind\nbvar}^{(\nbdim)}\right]\tilde{x}_{\nbind\nbvar}}{
\sum_{\nbind,\nbvar}\left(\mathbb{V}\left[\hat x_{\nbind\nbvar}^{(\nbdim)}\right]+\mathbb{E}\left[\hat x_{\nbind\nbvar}^{(\nbdim)}\right]^2\right)}.
\end{align*}
In the asymptotic framework where $\sigma^2$ tends to 0, \cite{Huet} showed that the expectation of $\hat x_{\nbind\nbvar}^{(\nbdim)}$ is equal to $\tilde x_{\nbind\nbvar}^{(\nbdim)}$. \cite{Verbanck13} approximated the variance of $\hat x_{\nbind\nbvar}^{(\nbdim)}$ by the noise variance $\frac{1}{\rkX}\sigma^2$.
Using these assumptions, \cite{Verbanck13} showed that the shrinkage terms can be written as the ratio between the variance of the signal and the total variance for the $\nbdim$ dimension, that they estimated using a plug-in estimator:
\begin{eqnarray}\label{estim_phi}
\Hat{\phi}_{\nbdim}=\frac{
\lambda_{\nbdim}-\frac{\Nbind \Nbvar}{\rkX}\hat{\sigma}^2}{
\lambda_{\nbdim}
} \text{ for all } \nbdim \text{ from 1 to } \Nbdim.
\end{eqnarray}
Although the theoretical properties of this estimator have not been exhibited, the simulation study conducted indicates that retaining this estimate for the shrinkage terms substantially reduces the mean squared error.

Thus, the regularized PCA solution $\hat{\bfX}^{rPCA}$ is defined by \cite{Verbanck13} as follows:
\begin{eqnarray}\label{acpreg}
\hat{x}^{rPCA}_{\nbind \nbvar}=\sum_{\nbdim =1}^{\Nbdim}\hat{\phi}_{\nbdim}\sqrt{\lambda_{\nbdim}}u_{\nbind \nbdim}v_{\nbvar \nbdim}.
\end{eqnarray}

\subsubsection{PCA on incomplete data\label{sec:pcaincomplete}}
With missing values, the classical solution to perform PCA is determined by minimizing the criterion $\parallel\Hat{\bfX}-\bfX\parallel^2$ on the observed data only. This is equivalent to introducing a weight matrix $\bfW$, where $w_{\nbind\nbvar}=0$ if $x_{\nbind\nbvar}$ is missing and $w_{\nbind\nbvar}=1$ otherwise, in the criterion which becomes $\parallel\bfW*\left(\Hat{\bfX}- \bfX\right)\parallel^2$ where $*$ is the Hadamard product.  To minimize this criterion, it is possible to use an EM algorithm called iterative PCA \citep{Kiers97}. The algorithm essentially sets the missing elements at initial values, performs the PCA on the completed data set, imputes the missing values with values predicted by the model (\ref{reconst_partial}) using a predefined number of dimensions ($\Nbdim$), and repeats the procedure on the newly obtained matrix until the total change in the matrix falls below an empirically determined threshold.
However such algorithms which alternate a step of estimation of the parameters using a singular value decomposition and a step of imputation of the missing values are known to suffer from overfitting problems. This means that the observed values are well fitted but the quality of prediction is poor. This occurs especially when the relationships between variables are low and/or when the number of missing values is high. To avoid these problems of overfitting, \citep{JosseHusson12} proposed to alternate the imputation and estimation steps by regularized PCA (\ref{acpreg}). The new algorithm is then called regularized iterative PCA.

Thus, the regularized iterative PCA algorithm can be used as a single imputation method since it produces a completed data set from the incomplete one. As stated in the introduction, performing multiple imputation requires taking into account the uncertainty of the estimation of the imputation model's parameters.
In this aim, we suggest a Bayesian approach to get $\Nbtab$ matrices $(\hat{\bfX}_{\nbtab})_{1 \leq \nbtab \leq \Nbtab}$  which will be obtained using draws from the posterior distribution of $\tilde{\bfX}$. Before describing the Bayesian approach on a data set with missing values, we present it on a complete data set.

\subsubsection{Bayesian PCA on complete data\label{bayesian_PCA}}
\citep{Verbanck13} proposed a Bayesian treatment of the PCA model using the following prior distribution for $\tilde{x}_{\nbind \nbvar}^{(\nbdim)}$:
\begin{eqnarray*}
\tilde{x}_{\nbind \nbvar}^{(\nbdim)}\sim \mathcal{N}(0,\tau_{\nbdim}^2)& \text{for all } 1\leq \nbdim\leq \Nbdim.
\end{eqnarray*}
Combining this prior distribution with the PCA model (\ref{pcamodel}), the posterior distribution has an explicit form: it is a normal distribution whose parameters depend on $\tau_{\nbdim}$ and $\sigma$:
\begin{align*}
p\left(\tilde x_{\nbind\nbvar}^{(\nbdim)}\vert x_{\nbind\nbvar}^{(\nbdim)}\right)=\mathcal{N}\left(\frac{\tau_{\nbdim}^2}{\tau_{\nbdim}^2+\frac{1}{\rkX}\sigma^2}x_{\nbind\nbvar}^{(\nbdim)},\frac{\tau_{\nbdim}^2\frac{\sigma^2}{\rkX}}{\tau_{\nbdim}^2+\frac{\sigma^2}{\rkX}}\right).
\end{align*}
Using an empirical Bayesian approach, $\tau_{\nbdim}$ and $\sigma$ are fixed from their estimates from the data as:
\begin{eqnarray*}
\hat{\tau}^2_{\nbdim}&=&\frac{1}{\Nbind\Nbvar}\lambda_{\nbdim}-\frac{\hat{\sigma}^2}{\rkX}
\end{eqnarray*} and $\hat{\sigma}^2$ defined in (\ref{estim_sigma}).
Thus, \citep{Verbanck13} showed that the posterior distribution of $\tilde{x}_{ij}^{(s)}$ is a normal distribution which has for expectation $\hat{x}^{(s)\,rPCA}_{ij}$ \eqref{acpreg} and for variance $\frac{{\hat{\sigma}}^2 {\hat{\phi}_s}}{\rkX}$ where $\phi_s$ given by $\frac{\tau_{\nbdim}^2}{\tau_{\nbdim}^2+\frac{\sigma^2}{\rkX}}$ is estimated by plug-in which corresponds to the estimate given in (\ref{estim_phi}).

Note that this modelling is in line with the one of \cite{Efron72} for a matrix $\tilde{\bfX}$ of full rank, and can be seen as a truncated version.
 \subsubsection{Bayesian PCA on incomplete data}
 \label{DA}
Generally, when a data set contains missing values, the posterior distribution of model parameters is often intractable. An algorithm which can be used in this context is the data augmentation (DA) algorithm \citep{Tanner87}. It consists in `augmenting' the observed data by predictions on missing data. The posterior becomes easier to calculate because the data set has become complete. DA simulates alternatively imputed values and parameters using a Markov chain which converges in probability to the observed posterior distribution. The algorithm consists of two steps:
\begin{itemize}
\item[(I)] imputing from the current parameters and the observed data,
\item[(P)] drawing of new parameters from the posterior given the new imputation and a prior distribution on the model's parameters.
\end{itemize}
Steps (I) and (P) are repeated a predefined number of times. At the end of the algorithm draws from the posterior distribution are obtained from an incomplete data set.\\

Inspired by the data augmentation algorithm to perform draws of $\tilde{x}$ in its posterior distribution, we essentially perform the two following steps:
\begin{itemize}
\item[(I)]given $\tilde{\bfX}$ and $\hat\sigma^2$, imputing the missing values $x_{\nbind\nbvar}$ by a draw from the predictive distribution $\mathcal{N}\left(\tilde{x}_{\nbind\nbvar},\hat\sigma^2\right)$
\item[(P)]drawing $\tilde{x}_{\nbind\nbvar}$ from its posterior distribution $\mathcal{N}\left(\hat{x}_{\nbind\nbvar}^{rPCA},\frac{{\hat{\sigma}}^2 \sum_s{\hat{\phi}_{\nbdim}}}{\rkX}\right)$ where $\hat{x}_{\nbind\nbvar}^{rPCA},\hat{\sigma}^2$ and $(\hat{\phi}_{\nbdim})_{1\leq\nbdim\leq \Nbdim}$ are calculated from the completed data set obtained from step (I).
\end{itemize}
Note that the estimates of $\phi$ and $\sigma$, that appear in the posterior distributions of $\tilde{\bfX}$, are updated by their maximum likelihood estimates in step (P), and are not fixed. 
Thus, it can be viewed as a marriage between a DA algorithm and an EM algorithm with unknown convergence properties.
\subsection{Multiple imputation with the BayesMIPCA algorithm}
\subsubsection{Presentation of the algorithm\label{mipca}}

In addition providing a posterior distribution of the parameters from an incomplete data set, the data augmentation algorithm can also be straightforwardly used to get multiple imputed data sets. 
To do so, after a burn-in step,  we simply keep $\Nbtab$ approximately independent draws leading to $\Nbtab$ imputed data sets. Thus, an imputed data set is saved at regular intervals.

This procedure of multiple imputation with Bayesian PCA is thus called the BayesMIPCA method. The details of the algorithm are as follows:
\begin{enumerate}
\item Initialization:
\begin{itemize}
\item calculate the matrix of means $\bfM^{[0]}$ which is the matrix of size $\Nbind\times \Nbvar$ with each row being the vector of the means of each column of the incomplete data set $\bfX$. The means are computed on the observed values.
\item centre $\bfX$: $\bfX^{[0]}\leftarrow \bfX-\bfM^{[0]}$. Since $\bfX$ is incomplete, $\bfX^{[0]}$ is also incomplete.
\item estimate the initial parameters $\tilde{\bfX}^{[0]}, \sigma^{2\,[0]}$ using, for instance, the regularized iterative PCA algorithm on $\bfX^{[0]}$
\end{itemize}
\item Burn in: for $\ell$ from 1 to \verb|Lstart|
\begin{itemize}
\item[(I)\textbullet] perform a random imputation according to the current parameters (drawn from the predictive distribution): $\bfX^{[\ell]}\leftarrow \bfW *\bfX ^{[\ell-1]}+(\bf{1}-\bfW)*(\tilde{\bfX}^{[\ell-1]}+ \bfE)$ where ${\bf {1}}_{I\times J}$ being a matrix with only ones and $\bfE_{\Nbind\times \Nbvar}=(\varepsilon_{\nbind\nbvar})_{1\leq\nbind\leq\Nbind ,1\leq\nbvar\leq\Nbvar}$ being a matrix of independent residuals so that $\varepsilon_{\nbind\nbvar}\sim \mathcal{N}(0,{{\hat{\sigma}}}^{2[\ell-1]})$; therefore $\bfX^{[\ell]}$ contains no missing values
\item add the matrix of means $\bfX^{[\ell]}\leftarrow\bfX^{[\ell]}+\bfM^{[\ell-1]}$
\end{itemize}
\begin{itemize}
\item[(P)\textbullet] calculate $\bfM^{[\ell]}$, the matrix of means of $\bfX^{[\ell]}$
\item centre the imputed data $\bfX^{[\ell]}\leftarrow\bfX^{[\ell]}-\bfM^{[\ell]}$
\item evaluate posterior parameters: calculate ${\hat{\bfX}}^{[\ell]}$, ${\hat{\sigma}}^{2\,[\ell]}$ and ${\hat{\phi}}^{[\ell]}$ from which we can deduce ${\hat{\bfX}}^{rPCA[\ell]}$
\item draw new parameters from the posterior: draw $\tilde{x}_{ij}^{[\ell]}$ from $\mathcal{N}\left(\hat{x}_{\nbind\nbvar}^{rPCA[\ell]},\frac{{\hat{\sigma}}^{2[\ell]} \sum_s{\hat{\phi}^{[\ell]}_{\nbdim}}}{\rkX}\right)$.
\end{itemize}
\item Create $\Nbtab$ imputed data sets: for $\nbtab$ from 1 to $\Nbtab$ alternate steps (I) and (P) \verb|L| times. \verb|L| is fixed and should be large enough to obtain independent imputations from one data set to another.
\end{enumerate}
\subsubsection{Modelling and analysis considerations}
The parameter $\Nbdim$ is supposed to be known \textit{a priori}. Many strategies are available in the literature to select a number of dimensions from a complete data set in PCA \citep{Jolliffe02}. Cross-validation \citep{Bro08} or an approximation of cross-validation such as generalized cross-validation \citep{Jossenbaxe11} perform well. We suggest these approaches since they can be directly extended to incomplete data \citep{JosseHusson12}. 

A simple chain is used to perform multiple imputation by data augmentation: \verb|Lstart| iterations are  passed in order to forget the dependence between the current settings and the initial parameters. \verb|Lstart| is equal to 1000 in our case. The $\Nbtab$ imputed data sets are obtained after \verb|Lstart+L|, \verb|Lstart+2*L|, \verb|Lstart+3*L|,\dots, \verb|Lstart+|$\Nbtab$\verb|*L| iterations with \verb|L| equal to 100.

Assessing the convergence of this kind of algorithm is still an open area of research. In practice, we investigate the values of some summaries, as sample moments or quantiles, through several iterations of the algorithm \citep{Schafer97}. The number of iterations required to observe stationarity for the summaries defines \verb|Lstart|, the number of iterations for the burn in step. Then, the autocorrelation of the summaries is investigated to determine a minimum value for \verb|L|.

Concerning the choice of $\Nbtab$, generating three to five data sets is usually enough in multiple imputation \citep{Rubin87}. However, due to increasing computational power, it is possible to generate a greater number of imputed data sets \citep[p.49]{VB12}. We use $\Nbtab=20$.

\subsection{Combining results from multiple imputed data sets}
As mentioned in the introduction, the aim of a multiple imputation procedure is to estimate a parameter and its variance from incomplete data. We detail hereafter the methodology described in \citep{Rubin87,Marshall09} to combine the results from multiple imputed data sets under the assumption of an estimator normally distributed and evaluated on a large sample. Note that this methodology is the same whatever the multiple imputation method used.
Let $\psi$ denote a quantity of interest that we want to estimate from an incomplete data set. 
To estimate this quantity and a confidence interval from $M$ imputed data sets obtained from a multiple imputation method, the following steps are performed:
\begin{itemize}
\item for $\nbtab=1,...\Nbtab$, $\hat\psi_\nbtab$ is computed on the imputed data set $\nbtab$ as well as its variance $\widehat{Var}(\hat\psi_\nbtab)$;
\item the results are pooled as: 
\begin{eqnarray*}\hat \psi&=& \frac{1}{\Nbtab}\sum_{\nbtab=1}^{\Nbtab}\hat{\psi}_\nbtab,\\
\widehat{Var}(\hat\psi)&=&\frac{1}{\Nbtab}
\sum_{\nbtab=1}^{\Nbtab}\widehat{Var}\left(\hat{\psi}_{\nbtab}\right)
+\left(1+\frac{1}{\Nbtab}\right)\frac{1}{\Nbtab-1}\sum_{\nbtab=1}^{\Nbtab}{\left(\hat{\psi}_{\nbtab}-\hat{\psi}\right)^2}.\end{eqnarray*}
The estimate of the variability of $\hat{\psi}$ is composed of two terms: the within-imputation variance corresponding to the sampling variability and the between-imputation variance corresponding to the variability due to missing values. The factor $(1 + \frac{1}{\Nbtab})$ corrects the fact that $\widehat{\psi}$ is an estimate for a finite number of imputed tables;
\item the 95\% confidence interval is calculated as:
\begin{eqnarray*}\hat{\psi}\pm t_{\nu,.975}\sqrt{\widehat{Var}(\hat\psi)}
\end{eqnarray*}
where $t_{\nu,.975}$ is the quantile corresponding to probability $.975$ of the Student's $t-$distribution with $\nu$ degrees of freedom estimated as suggested by \citep{Barnard99}.
\end{itemize}
\section{Evaluation of the methodology \label{SimulationNorm}}
To assess the multiple imputation method based on PCA, we conducted an extensive simulation study. We generated data sets drawn from normal distributions. These data sets differ with respect to the number of variables, the number of individuals and the strength of relationships between variables. We also considered real data sets.
The code to reproduce all the simulations with the R software \citep{Rsoft} is available on the webpage of the first author.

\subsection{Competing algorithms\label{comp_algo}}
The BayesMIPCA method is compared to the two following multiple imputation methods: a first one based on joint modelling implemented in the  R-package Amelia \citep{ameliapackage,Honaker11} and a second one based on chained equations implemented in the R-package mice \citep{micepackage,vanBuuren11}.
\begin{itemize}
\item \textbf{Amelia} imputes missing values by assuming a multivariate normal distribution for the variables. The uncertainty on the parameters is spread using a bootstrap approach \citep{Honaker10}. More precisely, $\Nbtab$ bootstrap incomplete data sets are generated and on each incomplete data set, the covariance matrix is estimated using an expectation-maximization algorithm. Then, the  $\Nbtab$ covariance matrices are used to produce $\Nbtab$ imputed data sets. 
The algorithm is implemented in the function \verb|amelia|. In the presence of high collinearity between variables, or a number of individuals too low compared to the number of variables, the variance-covariance matrix is not computationally invertible and therefore imputation under the normal distribution is not possible. In order to perform imputation in such conditions, it would be possible to introduce a ridge term to improve the conditioning of the regression problem.
\item \textbf{Mice (BayesMI method)} requires specifying a model for each variable with missing data.
The BayesMI method provides an imputation by regression for continuous variables where uncertainty on regression parameters is spread using a Bayesian approach. This method is implemented in the function \verb|mice.impute.norm| in the mice package. In the same way as the Amelia package, a ridge term could be introduced to overcome collinearity problems or lack of observations. It is also possible to specify  a conditional model where only a subset of variables is used as explanatory variables in each regression model.
\item\textbf{Listwise deletion} deletes individuals with missing values. This is not a multiple imputation method, but it is a benchmark for the variability of estimates. Because listwise deletion is equivalent to performing a statistical method on a sub-sample, variability should be greater than for a multiple imputation method.
\end{itemize}

\subsection{Simulation study with a block diagonal structure for the covariance matrix\label{simulationstudy}}
\subsubsection{Simulation design}
A data set $\bfX$ with $\Nbind$ rows and $\Nbvar$ columns is drawn from a normal distribution with null expectation and variance-covariance matrix of the form:

$$\left(
\begin{array}{cc}

\begin{array}{ccccc}
1&\rho&\ldots&\rho&\rho\\
\rho&1&\ldots&\rho&\rho\\
\vdots&\vdots&\ddots&\vdots&\vdots\\
\rho&\rho&\ldots&1&\rho\\
\rho&\rho&\ldots&\rho&1
\end{array}
& 0\\
0&
\begin{array}{ccc}
1&\ldots&\rho\\
\vdots&\ddots&\vdots\\
\rho&\ldots&1
\end{array}
\end{array}\right)$$
with $0<\rho<1$.
The variables are divided into two groups of size 2/3 and 1/3. Within each group, the pairwise correlation between variables is equal to $\rho$ and the two groups of correlated variables are independent. Thus, the number of underlying dimensions $\Nbdim$ is equal to 2.
The coefficient $\rho$ takes the values 0.9 or 0.3 to obtain strong or weak relationships between variables.
The number of variables is $\Nbvar = 6$ or $\Nbvar = 60$ and the number of individuals $\Nbind = 30$ or $\Nbind = 200$.
Then, we insert missing values (10\% or 30\%) completely at random, meaning that the probability that a value is missing is unrelated to the value itself and any values in the data set, missing or observed.
Each simulation is repeated $\Nbsim=1000$ times. 

Note that this simulation design is also suited for the competing algorithms, which are dedicated to normal data: the one in the Amelia package assumes multivariate normal distribution and the one in the mice package assumes a regression model for each variable.
\subsubsection{Criteria}

We consider three quantities of interest $\psi$ to be estimated from incomplete data: the expectation of a variable $\mathbb{E}[X_1]$, the correlation coefficient $\rho(X_{\Nbvar-1},X_{\Nbvar})$ between two variables and the regression coefficient $\beta_{X_2}$, which corresponds to the coefficient of the first explanatory variable in the regression model where $X_1$ is the response and $(X_2,\dots,X_{\Nbvar})$ the explanatory variables. The first quantity of interest is an indicator on a distribution of one variable and others on the relationships between variables. 

The criteria of interest are the bias $\frac{1}{\Nbsim}\sum_{\nbsim=1}^{\Nbsim}{\hat{\psi}_{\nbsim}-\psi}$, the root mean squared error (RMSE) $\sqrt{\frac{1}{\Nbsim}\sum_{\nbsim=1}^{\Nbsim}{(\hat{\psi}_{\nbsim}-\psi)^2}}$, the median (over the $\Nbsim$ simulations) of the confidence intervals width as well as the 95\% coverage. This latter is calculated as the percentage of cases where the true value $\psi$ is within the 95\% confidence interval. ``The 95\% coverage should be 95\% or higher. Coverages below 90\% are considered undesirable'' \citep[p.47]{VB12}.

As a benchmark, we also calculated the confidence intervals for the data sets without missing values which we call ``Full data''. The confidence interval obtained by multiple imputation should be greater.

\textbf{Remark.} Confidence intervals are based on the assumption that $\hat{\psi}$ is normally distributed. This is not true for the correlation coefficient $\rho$. Therefore a Fisher $z$ transformation is needed \citep{Schafer97}:\begin{eqnarray*}
z(\rho)=\frac{1}{2}\text{ln}\left(\frac{1+\rho}{1-\rho}\right)
\end{eqnarray*}

\subsubsection{Results\label{simu}}
For the point estimate of the expectation of a variable ($\psi=\mathbb{E}[X_1]$), all methods give good results: they produce unbiased estimates (results not shown here).  In addition, the root mean squared errors are of the same order of magnitude. Thus, the simulations do not highlight differences between the methods in terms of point estimate.  
Concerning the estimate of the variability of the estimator, Table \ref{mean} gives the median of the confidence intervals width and the 95\% coverage over the 1000 simulations for different simulations' configurations. In addition, when an algorithm fails on a configuration, no result is given. 
 \begin{table}[h!]
 \begin{center}
\caption{Results for the mean. Median confidence intervals width and 95\% coverage for $\psi=\mathbb{E}[X_1]$ estimated by several methods (Listwise deletion, Amelia, BayesMI and BayesMIPCA) for different configurations varying the number of individuals ($\Nbind=30$ or $200$), the number of variables ($\Nbvar=6$ or $60$), the strength of the relationships between variables ($\rho=0.3$ or $0.9$) and the percentage of missing values (10\% or 30\%). For each configuration, 1000 data sets with missing values are generated. Some values are not available because of failures of the algorithms.}{
$\begin{array}{c|c|c|c|c||c|c|c|c||c|c|c|c|}
\cline{2-13}
&\multicolumn{4}{c||}{\text{parameters}}&\multicolumn{4}{c||}{\text{confidence interval width}}&\multicolumn{4}{c|}{\text{coverage}}\\ \cline{2-13}
&\Nbind&\Nbvar&\rho&\text{\%}&\rotatebox{80}{\text{LD}}&\rotatebox{80}{\text{Amelia}}&\rotatebox{80}{\text{BayesMI}}&\rotatebox{80}{\text{BayesMIPCA}}&\rotatebox{80}{\text{LD}}&\rotatebox{80}{\text{Amelia}}&\rotatebox{80}{\text{BayesMI}}&\rotatebox{80}{\text{BayesMIPCA}}\\
\cline{2-13}
1&30&6&0.3&0.1&1.034&0.803&0.805&0.781&0.936&0.955&0.953&0.950\\
2&30&6&0.3&0.3&&&1.010&0.898&&&0.971&0.949\\
3&30&6&0.9&0.1&1.048&0.763&0.759&0.756&0.951&0.952&0.95&0.949\\
4&30&6&0.9&0.3&&&0.818&0.783&&&0.965&0.953\\
5&30&60&0.3&0.1&&&&0.775&&&&0.955\\
6&30&60&0.3&0.3&&&&0.864&&&&0.952\\
7&30&60&0.9&0.1&&&&0.742&&&&0.953\\
8&30&60&0.9&0.3&&&&0.759&&&&0.954\\
9&200&6&0.3&0.1&0.383&0.291&0.294&0.292&0.938&0.947&0.947&0.946\\
10&200&6&0.3&0.3&0.864&0.328&0.334&0.325&0.942&0.954&0.959&0.952\\
11&200&6&0.9&0.1&0.385&0.281&0.281&0.281&0.945&0.953&0.95&0.952\\
12&200&6&0.9&0.3&0.862&0.288&0.289&0.288&0.942&0.948&0.951&0.951\\
13&200&60&0.3&0.1&&&0.304&0.289&&&0.957&0.945\\
14&200&60&0.3&0.3&&&0.384&0.313&&&0.981&0.958\\
15&200&60&0.9&0.1&&&0.282&0.279&&&0.951&0.948\\
16&200&60&0.9&0.3&&&0.296&0.283&&&0.958&0.952\\
\cline{2-13}
\end{array}$
}
\label{mean}
\end{center}
\end{table}

With the current version of Amelia \citep{ameliapackage}, it is impossible to get results for the cases where $\Nbind<\Nbvar$ for our simulations.
These problems may be a pitfall of the implementation of the method since in theory using regularization may be able to handle such situations. Nevertheless, it would still be difficult to run the simulations since only expertise allows the selection of the tuning parameter in a missing data framework. For these reasons no results are provided for cases 5, 6, 7, 8. In addition, the algorithm regularly fails when there are many missing values. This problem is exacerbated when the number of variables is high or when the number of individuals is low (cases 2, 4, 13, 14, 15, 16).
Since the imputation by chained equations using the BayesMI method requires estimating the parameters of a regression model for each variable to be imputed, it suffers from the same kind of problems as the Amelia's algorithm. The solution to this problem consists in selecting a subset of explanatory variables for each conditional model. But it is difficult to make an appropriate selection of the predictors and there is no fully automatic default solution for the BayesMI method. For this reason no output is provided in the case where $\Nbind<\Nbvar$.
Finally, the listwise deletion cannot be performed on data sets where the rate of missing data is too high compared to the number of entries.
On the contrary, multiple imputation using the BayesMIPCA method can be applied on data sets of various kinds: when the collinearity between variables is weak or strong, when the rate of missing data is large or small, the number of individuals less than or greater than the number of variables.

All the algorithms give valid coverage, close to 95\% in all conditions where they perform. As expected, the confidence intervals for the multiple imputation methods are larger than those obtained from a complete dataset (0.734 for $\Nbind=30$ and 0.278 for $\Nbind=200$) and smaller than those obtained by listwise deletion. However the width of the confidence interval is often shorter for the BayesMIPCA method than for the other multiple imputation algorithms (particularly on the cases 1, 2, 4, 13, 14, 16).

Concerning the correlation coefficient, as for the expectation, the main differences between the algorithms are highlighted using the criteria that assess the variability of the estimator. Results are gathered in Table \ref{cor}. Note that according to the true value of $\rho$, the width of the confidence interval is not the same, because $\rho$ lies in the interval $[-1,1]$. If $\rho=0.9$, then $\rho$ is close to a bound and the interval is necessarily shorter than if $\rho=0.3$. For this reason, the widths of the confidence intervals have to be compared to those obtained from a complete data set.
Thus, the median width of the confidence intervals obtained from a complete data set is considered as the reference and the increase from this width is given in Table \ref{cor}.
 \begin{table}[ht]
\begin{center}
 \caption{Results for the correlation coefficient. Increase of the median of the widths of the confidence intervals obtained by the imputation method and the one obtained by full data as well as 95\% coverage for $\psi=\rho(X_{\Nbvar-1},X_\Nbvar)$. Results are given for several methods (Listwise deletion, Amelia, BayesMI and BayesMIPCA) on different configurations varying the number of individuals ($\Nbind=30$ or $200$), the number of variables ($\Nbvar=6$ or $60$), the strength of the relationships between variables ($\rho=0.3$ or $0.9$) and the percentage of missing values (10\% or 30\%). For each set of parameters, 1000 data sets with missing values are generated. Some values are not available because of failures of the algorithms.}{
$\begin{array}{c|c|c|c|c||c|c|c|c||c|c|c|c|} \cline{2-13}
&\multicolumn{4}{c||}{\text{parameters}}&\multicolumn{4}{c||}{\text{confidence interval width}}&\multicolumn{4}{c|}{\text{coverage}}\\ \cline{2-13}
&\Nbind&\Nbvar&\rho&\text{\%}&\rotatebox{80}{\text{LD}}&\rotatebox{80}{\text{Amelia}}&\rotatebox{80}{\text{BayesMI}}&\rotatebox{80}{\text{BayesMIPCA}}&\rotatebox{80}{\text{LD}}&\rotatebox{80}{\text{Amelia}}&\rotatebox{80}{\text{BayesMI}}&\rotatebox{80}{\text{BayesMIPCA}}\\
\cline{2-13}
1 & 30 & 6 & 0.3 & 0.1 &+36\%&+16\%&+17\%&+14\%&0.938 &0.957 &0.964 &0.963\\
2 & 30 & 6 & 0.3 & 0.3 &  &  &+56\%&+36\%& & &0.976 &0.956\\
3 & 30 & 6 & 0.9 & 0.1 &+49\%&+32\%&+31\%&+14\%&0.935 &0.968 &0.962 &0.968\\
4 & 30 & 6 & 0.9 & 0.3 &  &  &+221\%&+40\%& & &0.974 &0.983\\
5 & 30 & 60 & 0.3 & 0.1 &  &  &  &+13\%& & & &0.971\\
6 & 30 & 60 & 0.3 & 0.3 &  &  &  &+27\%& & & &0.989\\
7 & 30 & 60 & 0.9 & 0.1 &  &  &  &+13\%& & & &0.976\\
8 & 30 & 60 & 0.9 & 0.3 &  &  &  &+26\%& & & &0.99\\
9 & 200 & 6 & 0.3 & 0.1 &+38\%&+11\%&+12\%&+10\%&0.959 &0.947 &0.952 &0.967\\
10 & 200 & 6 & 0.3 & 0.3 &+202\%&+45\%&+47\%&+27\%&0.939 &0.942 &0.949 &0.974\\
11 & 200 & 6 & 0.9 & 0.1 &+40\%&+8\%&+9\%&+6\%&0.958 &0.953 &0.956 &0.967\\
12 & 200 & 6 & 0.9 & 0.3 &+247\%&+30\%&+43\%&+23\%&0.940 &0.948 &0.943 &0.973\\
13 & 200 & 60 & 0.3 & 0.1 &  &  &+15\%&+8\%& & &0.964 &0.981\\
14 & 200 & 60 & 0.3 & 0.3 &  &  &+55\%&+21\%& & &0.945 &0.989\\
15 & 200 & 60 & 0.9 & 0.1 &  &  &+23\%&+6\%& & &0.914 &0.969\\
16 & 200 & 60 & 0.9 & 0.3 &  &  &+83\%&+13\%& & &0.683 &0.985\\
\cline{2-13}
\end{array}$
}
\label{cor}
\end{center}
\end{table}
The BayesMI and Amelia methods produce confidence intervals of similar widths while they are shorter with the BayesMIPCA method which moreover has a better coverage.
This good behaviour of the BayesMIPCA method can be explained by the properties of the imputation model. Indeed, PCA is a dimensionality reduction method used to isolate the relevant information of a data set. This makes it very stable and implies that the imputation from a table to another does not change much: the between-variability is lower than for the other methods, which explains that the confidence intervals are shorter. 
When the strength of the relationships between variables is low (cases 2, 10, 14), the difference between the width of the confidence intervals obtained from the BayesMIPCA method and the width of those obtained from the two other methods is moderate. At the most the increase between the median of the widths of the confidence intervals and the median of the widths obtained from a complete data set attempts $+55$\% for the BayesMI method versus $+21$\% for the BayesMIPCA one. However, when the relationships between variables are strong (cases 4, 12, 16), the BayesMI and Amelia algorithms encounter great difficulties. The width of the confidence interval obtained with BayesMI is up to 3 times larger than the one obtained from a complete set (case 4) versus 1.4 for the BayesMIPCA method. For the 16$^{th}$ case, it even leads to very bad results with a coverage close to 68\%.

The results on the estimate of the regression coefficient lead to the same conclusions as those already mentioned for the expectation and for the correlation coefficient: with BayesMIPCA, confidence intervals are shorter and coverages are accurate. In addition, the BayesMIPCA method systematically gives the smallest mean squared error. The results for this quantity are presented in the appendix.

\subsection{Simulation study with a fuzzy principal component structure\label{fuzzy}}

As a complement to the previous simulations in Section \ref{simulationstudy}, we assess the BayesMIPCA algorithm when the low dimensional structure of the data is less obvious. Instead of generating the data sets using covariance matrices with a two block diagonal structure, we generate covariance matrices at random as in \cite{Joe06}. More precisely, the draw is uniform over the space of positive definite correlation matrices. The method is implemented in the R package clusterGeneration \citep{clusterGeneration}.
We generated two covariance matrices, one for $\Nbvar=6$ variables and another one for $\Nbvar=60$ variables. From each matrix, $\Nbsim=1000$ data sets are drawn varying the number of individuals ($\Nbind=30$ or $\Nbind=200$) and the percentage of missing values (10\% or 30\%). Multiple imputation (using $\Nbtab=20$ imputed data sets) is performed on each of them to estimate the quantities of interest (an expectation, a regression coefficient and a correlation coefficient). The quality of the imputation is assessed using the same quantities of interest and the same criteria as those used in Section \ref{simulationstudy}. The results for the mean are gathered in Table \ref{table5} and the ones for the correlation coefficient are gathered in Table \ref{table6}.

Since the dimensional structure of the data is less obvious, the potential number of underlying dimensions is unknown \textit{a priori}. Thus, we are in a setting close to what happens with real data and we use cross-validation \citep{Bro08} to select $\Nbdim$, the number of underlying dimensions used in the BayesMIPCA algorithm. However, cross-validation is time consuming, consequently we cannot perform it for each configuration (\textit{i.e.} for a number of individuals, a number of variables and a percentage of missing values) and for each of the $\Nbsim=1000$ incomplete data sets. For this reason, for each configuration, the choice of $\Nbdim$ is based on cross-validation performed on 20 incomplete data sets only. This is sufficiently large because of the relative stability of the results. The most frequent number of underlying dimensions over the 20 simulations is retained.

 \begin{table}[h!]
\begin{center}
 \caption{Results for the mean. Median confidence intervals width and 95\% coverage for $\psi=\mathbb{E}[X_1]$ estimated by several methods (Listwise deletion, Amelia, BayesMI and BayesMIPCA) for different configurations varying the number of individuals ($\Nbind=30$ or $200$), the number of variables ($\Nbvar=6$ or $60$) and the percentage of missing values (10\% or 30\%). The data sets are drawn from a random covariance matrix. The number of underlying dimensions $\Nbdim$ is estimated by cross-validation. For each configuration, 1000 data sets with missing values are generated. Some values are not available because of failures of the algorithms.}{
$\begin{array}{c|c|c|c|c||c|c|c|c||c|c|c|c|}
\cline{2-13}
&\multicolumn{4}{c||}{\text{parameters}}&\multicolumn{4}{c||}{\text{confidence interval width}}&\multicolumn{4}{c|}{\text{coverage}}\\ \cline{2-13}
&\Nbind&\Nbvar&\text{\%}&\Nbdim&\rotatebox{80}{\text{LD}}&\rotatebox{80}{\text{Amelia}}&\rotatebox{80}{\text{BayesMI}}&\rotatebox{80}{\text{BayesMIPCA}}&\rotatebox{80}{\text{LD}}&\rotatebox{80}{\text{Amelia}}&\rotatebox{80}{\text{BayesMI}}&\rotatebox{80}{\text{BayesMIPCA}}\\ \cline{2-13}
 1&30&6&0.1&4&1.026&0.777&0.777&0.765&0.949&0.948&0.947&0.947\\
 2&30&6&0.3&2&&&0.945&0.839&&&0.965&0.948\\
 3&30&60&0.1&5&&&&0.786&&&&0.956\\
 4&30&60&0.3&5&&&&0.92&&&&0.956\\
 5&200&6&0.1&4&0.391&0.285&0.286&0.284&0.947&0.94&0.942&0.937\\
 6&200&6&0.3&4&&0.312&0.315&0.303&&0.945&0.954&0.937\\
 7&200&60&0.1&5&&&0.284&0.291&&&0.941&0.943\\
 8&200&60&0.3&5&&&0.359&0.321&&&0.971&0.941\\
\cline{2-13}
\end{array}$
}
\label{table5}
\end{center}
\end{table}
 \begin{table}[h!]
\begin{center}
 \caption{Results for the correlation coefficient. Increase of the median of the widths of the confidence intervals obtained by the imputation method and the one obtained by the full data, as well as 95\% coverage for $\psi=\rho(X_{\Nbvar-1},X_\Nbvar)$. Results are given for several methods (Listwise deletion, Amelia, BayesMI and BayesMIPCA) on different configurations varying the number of individuals ($\Nbind=30$ or $200$), the number of variables ($\Nbvar=6$ or $60$) and the percentage of missing values (10\% or 30\%). The data sets are drawn from a random covariance matrix. The number of underlying dimensions $\Nbdim$ is estimated by cross-validation. For each configuration, 1000 data sets with missing values are generated. Some values are not available because of failures of the algorithms.}{
$\begin{array}{c|c|c|c|c||c|c|c|c||c|c|c|c|}
\cline{2-13}
&\multicolumn{4}{c||}{\text{parameters}}&\multicolumn{4}{c||}{\text{confidence interval width}}&\multicolumn{4}{c|}{\text{coverage}}\\ \cline{2-13}
&\Nbind&\Nbvar&\text{\%}&\Nbdim&\rotatebox{80}{\text{LD}}&\rotatebox{80}{\text{Amelia}}&\rotatebox{80}{\text{BayesMI}}&\rotatebox{80}{\text{BayesMIPCA}}&\rotatebox{80}{\text{LD}}&\rotatebox{80}{\text{Amelia}}&\rotatebox{80}{\text{BayesMI}}&\rotatebox{80}{\text{BayesMIPCA}}\\
\cline{2-13}
1&30&6&0.1&4&+47\%&+10\%&+11\%&+30\%&0.944&0.959&0.962&0.947\\
2&30&6&0.3&2&&&+66\%&+70\%&&&0.977&0.911\\
3&30&60&0.1&5&&&&+10\%&&&&0.975\\
4&30&60&0.3&5&&&&+26\%&&&&0.991\\
5&200&6&0.1&4&+41\%&+3\%&+3\%&+9\%&0.946&0.953&0.953&0.954\\
6&200&6&0.3&4&&+14\%&+21\%&+31\%&&0.954&0.961&0.92\\
7&200&60&0.1&5&&&+4\%&+8\%&&&0.959&0.958\\
8&200&60&0.3&5&&&+39\%&+23\%&&&0.988&0.96\\
\cline{2-13}
\end{array}$
}
\label{table6}
\end{center}
\end{table}

The results for the mean are very similar to the ones obtained in Section \ref{simu}: the estimator is unbiased for all cases (results not shown here), the coverages are valid and the confidence intervals are shorter for the BayesMIPCA algorithm than for the others. On the contrary, the results for the correlation coefficient highlight the difficulties encountered by BayesMIPCA for data sets with a fuzzy principal component structure. In the cases 3, 4, 7 and 8, where the number of variables is high compared to the number of underlying dimensions estimated (\textit{cf.} Table \ref{table6}), the coverages are very good and the confidence interval widths are close to the ones obtained by the BayesMI method when it provides results. The hypothesis of an underlying signal in a lower dimensional space is likely in these cases, and consequently, the results are similar to those obtained with a two block structure for the covariance matrix. In the other cases, where the number of variables is small compared to the number of underlying dimensions estimated, the coverages remain satisfactory (greater than 90\%) but sometimes worse than previously: in cases 2 and 6 the coverage is close to 92\% instead of 95\%. Thus, the BayesMIPCA method is all the more efficient in the case of a low dimensional structure.\\

In order to go deeper and to deal with larger data, another configuration with 1000 individuals, 200 variables and 10\% of missing values is considered. The covariance matrix of size $200\times 200$ is drawn at random \citep{Joe06}. In this configuration, the cross-validation method does not provide a reliable number of dimensions (it gives as a solution the number of variables). Consequently, $\Nbdim=17$ dimensions are kept using an ad hoc strategy (by looking at the barplot of the eigenvalues). 
Because dealing with a big data set is time consuming, multiple imputation using only $\Nbtab=5$ imputed data sets is performed. The results for the BayesMI and the BayesMIPCA methods are gathered in Table \ref{table7} (the Amelia's algorithm failed on these simulations).

\begin{table}[h!]
\begin{center}
\caption{Results for the mean and the correlation coefficient. Bias, root mean squared error, median confidence intervals width and 95\% coverage for $\psi=\mathbb{E}[X_1]$ and $\psi=\rho(X_{\Nbvar-1},X_{\Nbvar})$ estimated by BayesMI and BayesMIPCA for a configuration with $\Nbind=1000$ individuals, $\Nbvar=200$ variables and 10\% of missing values. The data sets are drawn from a random covariance matrix. 1000 data sets with missing values are generated. Results for the full data are also provided.}{

\begin{tabular}{p{2cm}|c|c|c||c|c|}
\cline{2-6} &\multicolumn{3}{c||}{mean}&\multicolumn{2}{c|}{correlation coefficient}\\
\cline{2-6}     & BayesMI & BayesMIPCA& Full data&BayesMI & BayesMIPCA\\ \hline 
\multicolumn{1}{|l|}{bias} &-0.001 &0 &       0  &     0   &   0.011 \\ \hline
\multicolumn{1}{|l|}{rmse}  &0.032&  0.034 &   0.032&   0.032&  0.035\\ \hline
\multicolumn{1}{|l|}{confidence interval width}  &0.127 & 0.131 &   0.124  & +3.31\%& +9.09\%  \\ \hline \multicolumn{1}{|l|}{coverage}&0.955&  0.958   & 0.96 &   0.949 & 0.933 \\ \hline
\end{tabular}
}
\label{table7}
\end{center}
\end{table}

As previously, the coverages are greater than 90\% for the BayesMIPCA method, but nevertheless below 95\% for the correlation coefficient. The number of underlying dimensions is crudely approximated and we can suppose that in reality it is not sufficiently small compared to the number of variables to reach a coverage of 95\%. BayesMIPCA performs better in the case of a low dimensional structure.\\

Finally, some simulations are performed based on a real large data set. Therefore the low dimensional structure of the data set is again unclear. This data set is a subset of the million song dataset (MSD)\citep{MSD}. It contains 463715 songs (rows) and 90 acoustic features (variables) dealing with the timbre of the song. Each feature corresponds to a particular ``segment'', which is generally delimited by note onsets, or other discontinuities in the signal. It contains also a variable corresponding to the year of the song. The aim is to predict the year of a song using its features. In fact, listeners often have particular affection for music from certain periods of their lives, thus the predicted year could be useful as a basis for recommendation \citep{MSD}. This subset is available on the web page \verb|http://archive.ics.uci.edu/ml/datasets/YearPredictionMSD|.

To perform simulations from a real data set, the data set is preliminarily scaled to be more likely in lines with the assumption of a homoscedastic noise as stated in \ref{pcamodel}. We consider that this data set defines the population. Thus, the true value of the quantity of interest is known.
Here, we are interested in the regression coefficient corresponding to the first explanatory variable in the regression model predicting the year of the song.
To assess the multiple imputation methods, $\Nbsim=1000$ samples of size $\Nbind=300$ are drawn from the population, 10\% of missing values are added and multiple imputation is performed using $\Nbtab=20$ imputed data sets. The cross-validation procedure indicates that 8 dimensions should be retained. The results for the BayesMI method and the BayesMIPCA one are gathered in Table \ref{table8} (the Amelia's algorithm fails again which seems to be strongly related to the current version of their implementation).

\begin{table}[h!]
\begin{center}
\caption{Results for the regression coefficient. Bias, root mean squared error, median confidence intervals width and 95\% coverage for $\psi=\beta_{X_2}$ estimated by BayesMI and BayesMIPCA on a subset of size $463715\times 90$ of the million song dataset. Multiple imputation is performed on 1000 samples of size $\Nbind=300$, drawn from the population and become incomplete with 10\% of missing values.}{
\begin{tabular}{p{2cm}|c|c|c|}
\cline{2-4}     & BayesMI & BayesMIPCA& Full data\\ \hline 
\multicolumn{1}{|l|}{ bias} &0.112& -0.121  & 0.071 \\ \hline
\multicolumn{1}{|l|}{rmse}  &0.216& 0.152  &  0.148 \\ \hline
\multicolumn{1}{|l|}{confidence interval width}  &0.754& 0.479&    0.438 \\ \hline
\multicolumn{1}{|l|}{coverage}  & 0.911 &0.887 &   0.859 \\ \hline
\end{tabular}
}
\label{table8}
\end{center}
\end{table}

The BayesMIPCA method provides results close to the ones obtained from the complete data set. The under-coverage observed on the full data could be explained by the small size of the samples compared to the size of the population (300 vs 463715), also by the heterogeneity of the population. The sample size was selected in order to perform simulations in a reasonable time. BayesMIPCA provides results that are more convincing than those of BayesMI (smaller size of the confidence interval). We can suppose that on this real data set, the hypothesis of an underlying signal of lower dimension is likely, and the BayesMIPCA method is well suited.
\subsection{Simulations from real data}
Finally, in order to evaluate the method in practical situations, we perform simulations using four real data sets. In comparison to the previous ones (Section \ref{fuzzy}), here we do not sample from these data sets but consider them as real data sets: it means that each data set is a sample from an unknown population.
The first data set refers to $\Nbind=41$ athletes' performances during a decathlon event \citep{Facto}. It contains $\Nbvar=11$ variables, the trials plus the score obtained by the athletes which is strongly related to the 10 other variables. The second data set concerns an isoprenoid gene network in A.~Thaliana \citep{Wille04}. This gene network includes $\Nbvar= 39$ genes each with $\Nbind = 118$ gene expression profiles corresponding to different experimental conditions. The genetic data are known to present complex relationships. The third data set deals with $\Nbind=112$ daily measurements of $\Nbvar = 11$ meteorological variables and ozone concentration recorded in Rennes (France) during summer 2001 \citep{Cornillonenglish12}. The last data set comes from a sensory study \citep{Facto} where $\Nbind=21$ wines of Val de Loire were evaluated on $\Nbvar=29$ descriptors. The number of individuals is less than the number of variables for this data.

On each data set, 30\% of missing values is randomly added and the three multiple imputation methods (Sections \ref{mipca} and \ref{comp_algo}) are performed. The listwise deletion method cannot be used for this percentage of missing values. We repeat this process 1000 times.
As for the simulations (Section \ref{simulationstudy}), we focus on the following quantities: a mean $\mu$, a regression coefficient $\beta$, as well as a correlation coefficient $\rho$. Because we deal with true data sets, the true values for the quantities of interest are unknown. Indeed, these real data sets are samples from a larger unknown population. In Table \ref{simu_real}, we report the point estimate and the confidence interval for each quantity, as well as the ones obtained from the completed data sets.

The behaviour of the BayesMIPCA method is quite similar to the one observed on simulations: the method can be applied whatever the data set, and gives the smallest confidence interval. For many cases, the three multiple imputation methods provide similar results close to the ones obtained from the completed data sets. However, the BayesMI method seems very unstable on the data set Decathlon. For example, the median confidence interval width for the $\beta$ coefficient is equal to 3.363. This could be explained by the collinearity in the data set combined with a small number of individuals.
 \begin{table}[h!]
\begin{center}
 \caption{Mean of the point estimates and median confidence intervals width (or relative increase compared to the complete case) for $\mu$, $\rho$, $\beta$ over 1000 simulations. Results are given for several methods (Amelia, BayesMI and BayesMIPCA) on different real datasets (Decathlon, Isoprenoid, Ozone, Wine) with 30\% of missing values. Results for the full data are also provided. Some values are not available because of failures of the algorithms.}{
$\begin{array}{|c|c|c|c|c|c||c|c|c|c|}
\cline{3-10}
\multicolumn{2}{c|}{}&\multicolumn{4}{c||}{\text{estimate}}&\multicolumn{4}{c|}{\text{confidence interval width}}\\ \cline{3-10}
\multicolumn{2}{c|}{}&\rotatebox{80}{\text{Amelia}}&\rotatebox{80}{\text{BayesMI}}&\rotatebox{80}{\text{BayesMIPCA}}&\rotatebox{80}{\text{Full data}}&\rotatebox{80}{\text{Amelia}}&\rotatebox{80}{\text{BayesMI}}&\rotatebox{80}{\text{BayesMIPCA}}&\rotatebox{80}{\text{Full data}}\\
\hline
\multirow{4}{0.25cm}{$\mu$}&\text{Decathlon}&&0&0&0&&0.704&0.717&0.631\\ \cline{2-10}
&\text{Isoprenoid}&&0.003&0.004&0&&0.448&0.406&0.365\\ \cline{2-10}
&\text{Ozone}&0.002&0.002&0.001&0&0.403&0.409&0.402&0.374\\ \cline{2-10}
&\text{Wine}&&&0.014&0&&&0.998&0.91\\ \hline  \hline
\multirow{4}{0.6cm}{\begin{tabular}{c}$\rho$\end{tabular}}&\text{Decathlon}&&0.491&0.545&0.616&&+92\%&+47\%&0.396\\ \cline{2-10}
&\text{Isoprenoid}&&0.609&0.637&0.705&&+82\%&+44\%&0.185\\ \cline{2-10}
&\text{Ozone}&0.65&0.66&0.654&0.685&+38\%&+43\%&+30\%&0.2\\ \cline{2-10}
&\text{Wine}&&&0.536&0.607&&&+35\%&0.585\\ \hline \hline
\multirow{4}{0.6cm}{\begin{tabular}{c}$\beta$\end{tabular}}&\text{Decathlon}&&-0.149&-0.16&-0.175&&3.363&0.793&0.01\\ \cline{2-10}
&\text{Isoprenoid}&&0.134&0.076&0.203&&0.584&0.44&0.382\\ \cline{2-10}
&\text{Ozone}&0.423&0.42&0.408&0.409&0.4&0.43&0.412&0.273\\ \cline{2-10}
&\text{Wine}&&&0.841&0.949&&&0.746&0.302\\\hline 
\end{array}$
}
\label{simu_real}
\end{center}
\end{table}
\section{Conclusion}

Multiple imputation by Bayesian PCA provides valid confidence intervals for both quantities related to the marginal distribution of a variable as well as for quantities related to the relationships between variables from an incomplete continuous data set. Compared to its competitors, it often gives confidence intervals with a smaller width. This is due to the imputation based on PCA. Indeed, PCA is a dimensionality reduction method which isolate the relevant information from the noise. This makes the imputation stable and consequently decreases the variability of the estimator. In addition, the multiple imputation by Bayesian PCA can be easily performed on any kind of data where for instance the number of individuals is less than the number of variables, which is a configuration where other methods encounter difficulties. We have shown that the method is well suited when the hypothesis of an underlying signal of low dimension is verified. In practice, this hypothesis is often true for many data sets. Nevertheless, when the hypothesis of a structure of low dimension is not met, the BayesMIPCA method remains competitive. Note also that since the imputation is based on PCA, it is particularly well fitted to situations where the relationships between variables are linear, and more generally when the data can be considered as being generated from a PCA model.
Thus, the multiple imputation method BayesMIPCA has many advantages and is a flexible alternative to the classical multiple imputation procedures suggested in the literature. However, this method requires tuning a parameter which is the number of dimensions $\Nbdim$. We suggest the use of cross-validation, or of an approximation of cross-validation, such as generalized cross-validation described in \citep{Jossenbaxe11} to choose $\Nbdim$. Simulations not presented here indicated that the method is fairly robust to a misspecified choice for $\Nbdim$, as long as  $\Nbdim$ is not too small (to be able to capture the relevant information). The BayesMIPCA method is available as an R function on the webpage of the first author.

Future research includes the assessment of the suggested method in cases where there are complex interactions or relationships between variables or cases where for instance a variable $X_1$ and its squared $X_1^2$ are of interest. \citep{Seaman12} compared different strategies to handle this latter situation such as the JAV (just another variable) approach which considers the squared version as a new variable in itself without taking into account its link with $X_1$. \citep{Bartlett13} suggested another MI method to handle such situations better but it does not give the possibility to deal with missing values in all the variables in its current form.

The encouraging results of the Bayesian PCA for continuous variables prompt the extension of the method to perform multiple imputation for categorical variables using multiple correspondence analysis \citep{Green06} and using factorial analysis for mixed data \citep{Kiers91,pages14}.
\citep{Audigier13} suggested single imputation methods based on principal component methods for data with continuous, categorical and mixed variables showing good results to predict the missing entries.
However, the extension to multiple imputation is not straightforward, because the method presented for continuous variables is based on a Bayesian treatment of a joint model for all variables. The model is well known for PCA, but the model is yet unknown for multiple correspondence analysis and \textit{a fortiori} for the factor analysis of mixed data.
Further research would be required if a Bayesian approach of these principal component methods, and therefore multiple imputation based on these methods, was being considered.

\bibliographystyle{gSCS}
\bibliography{biblioarticle}

\begin{thebibliography}{10}
\providecommand{\url}[1]{\normalfont{#1}}
\providecommand{\urlprefix}{Available from: }

\bibitem{Dempster77}
{Dempster}~AP, {Laird}~NM, {Rubin}~DB. Maximum likelihood from incomplete data
  via the em algorithm. Journal of the Royal Statistical Society B.
  1977;\hspace{0pt}39:1--38.

\bibitem{Meng91}
{Meng}~XL, {Rubin}~DB. Using {EM} to {O}btain {A}symptotic
  {V}ariance-{C}ovariance {M}atrices: {T}he {SEM} {A}lgorithm. Journal of the
  American Statistical Association. 1991 Dec;\hspace{0pt}86(416):899--909.

\bibitem{Rubin87}
{Rubin}~DB. Multiple imputation for non-response in survey. Wiley; 1987.

\bibitem{Little02}
{Little}~RJA, {Rubin}~DB. Statistical analysis with missing data. New-York:
  Wiley series in probability and statistics; 1987, 2002.

\bibitem{Schafer97}
{Schafer}~JL. Analysis of incomplete multivariate data. London: Chapman \&
  Hall/CRC; 1997.

\bibitem{vanBuuren06}
{Van Buuren}~S, {Brand}~JPL, {Groothuis-Oudshoorn}~CGM, {Rubin}~DB. {Fully
  conditional specification in multivariate imputation}. Journal of Statistical
  Computation and Simulation. 2006;\hspace{0pt}76:1049--1064.

\bibitem{Besag74}
{Besag}~J. {Spatial Interaction and the Statistical Analysis of Lattice
  Systems}. Journal of the Royal Statistical Society Series B (Methodological).
  1974;\hspace{0pt}36(2).

\bibitem{liu14}
Liu~J, Gelman~A, Hill~J, Su~YS, Kropko~J. {On the stationary distribution of
  iterative imputations}. Biometrika. 2014 Mar;\hspace{0pt}:155--173.

\bibitem{Kropko13}
{Kropko}~J, {Goodrich}~B, {Gelman}~A, {Hill}~J. Multiple imputation for
  continuous and categorical data: Comparing joint and conditional approaches.
  Political Analysis. 2014;\hspace{0pt}.

\bibitem{JosseHusson12}
{Josse}~J, {Husson}~F. Handling missing values in exploratory multivariate data
  analysis methods. Journal de la Soci\'et\'e Fran\c caise de Statistique.
  2012;\hspace{0pt}153 (2):1--21.

\bibitem{Caussinus86}
{Caussinus}~H. Models and uses of principal component analysis (with
  discussion). In: Multidimensional data analysis. DSWO Press; 1986. p.
  149--178.

\bibitem{Candes09}
{Cand\`{e}s}~EJ, {Tao}~T. The power of convex relaxation: Near-optimal matrix
  completion. IEEE Trans Inf Theor. 2009 May;\hspace{0pt}56(5):2053--2080.

\bibitem{shabalin13}
Shabalin~A, Nobel~B. Reconstruction of a low-rank matrix in the presence of
  gaussian noise. Journal of Multivariate Analysis. 2013;\hspace{0pt}118(0):67
  -- 76.

\bibitem{Verbanck13}
{Verbanck}~M, {Josse}~J, {Husson}~F. Regularised {PCA} to denoise and visualise
  data. Statistics and Computing. 2013;\hspace{0pt}:1--16.

\bibitem{Josse15}
Josse~J, Sardy~S. Adaptive shrinkage of singular values. Statistics and
  Computing. 2015;\hspace{0pt}:1--10.

\bibitem{Huet}
Huet~S, Denis~J, Adamczyk~K. Bootstrap confidence intervals in nonlinear
  regression models when the number of observations is fixed and the variance
  tends to o. application to biadditive models. Statistics.
  1999;\hspace{0pt}32:203--227.

\bibitem{Kiers97}
{Kiers}~HAL. Weighted least squares fitting using ordinary least squares
  algorithms. Psychometrika. 1997;\hspace{0pt}62:251--266.

\bibitem{Efron72}
Efron~B, Morris~C. {Empirical Bayes on Vector Observations: An Extension of
  Stein's Method}. Biometrika. 1972;\hspace{0pt}59(2):335--347.

\bibitem{Tanner87}
{Tanner}~MA, {Wong}~WH. The calculation of posterior distributions by data
  augmentation. Journal of the American Statistical Association.
  1987;\hspace{0pt}82:805--811.

\bibitem{Jolliffe02}
{Jolliffe}~IT. Principal component analysis. Springer; 2002.

\bibitem{Bro08}
{Bro}~R, {Kjeldahl}~K, {Smilde}~AK, {Kiers}~HAL. Cross-validation of component
  model: a critical look at current methods. Anal Bioanal Chem.
  2008;\hspace{0pt}390:1241--1251.

\bibitem{Jossenbaxe11}
{Josse}~J, {Husson}~F. Selecting the number of components in \textsc{PCA} using
  cross-validation approximations. Computational Statististics and Data
  Analysis. 2011;\hspace{0pt}56(6):1869--1879.

\bibitem{VB12}
{Van Buuren}~S. Flexible imputation of missing data (chapman \& hall/crc
  interdisciplinary statistics). 1st ed. Chapman and Hall/CRC; 2012.

\bibitem{Marshall09}
Marshall~A, Altman~DG, Holder~RL, Royston~P. Combining estimates of interest in
  prognostic modelling studies after multiple imputation: current practice and
  guidelines. Bmc Medical Research Methodology. 2009;\hspace{0pt}9(5):57.

\bibitem{Barnard99}
{Barnard}~J, {Rubin}~DB. {S}mall {S}ample {D}egrees of {F}reedom with
  {M}ultiple {I}mputation. Biometrika. 1999;\hspace{0pt}86:948--955.

\bibitem{Rsoft}
{R Core Team}. R: A language and environment for statistical computing. R
  Foundation for Statistical Computing; Vienna, Austria. 2014;
  \urlprefix\url{http://www.R-project.org}.

\bibitem{ameliapackage}
{Honaker}~J, {King}~G, {Blackwell}~M. Amelia ii: A program for missing data.
  2014; r package version 1.7.2.

\bibitem{Honaker11}
{Honaker}~J, {King}~G, {Blackwell}~M. {Amelia II}: A program for missing data.
  Journal of Statistical Software. 2011;\hspace{0pt}45(7):1--47.

\bibitem{micepackage}
{Van Buuren}~S. mice. 2014; r package version 2.18.

\bibitem{vanBuuren11}
{Van Buuren}~S, {Groothuis-Oudshoorn}~CGM. {mice}: Multivariate imputation by
  chained equations in \textsc{R}. Journal of Statistical Software.
  2011;\hspace{0pt}45(3):1--67.

\bibitem{Honaker10}
{Honaker}~J, {King}~G. What to do about missing values in time series
  cross-section data. American Journal of Political Science.
  2010;\hspace{0pt}54:561--581.

\bibitem{Joe06}
Joe~H. Generating random correlation matrices based on partial correlations. J
  Multivar Anal. 2006 Nov;\hspace{0pt}97(10):2177--2189.

\bibitem{clusterGeneration}
Qiu~W, Joe~H. clustergeneration: random cluster generation (with specified
  degree of separation). 2013; r package version 1.3.1.

\bibitem{MSD}
Bertin-Mahieux~T, Ellis~D, Whitman~B, Lamere~P. The million song dataset. In:
  {Proceedings of the 12th International Conference on Music Information
  Retrieval ({ISMIR} 2011)}; 2011.

\bibitem{Facto}
{Husson}~F, {Josse}~J, {Le}~S, {Mazet}~J. Factominer: Multivariate exploratory
  data analysis and data mining with r. 2013; r package version 1.25;
  \urlprefix\url{http://CRAN.R-project.org/package=FactoMineR}.

\bibitem{Wille04}
{Wille}~A, {Zimmermann}~P, {Vranova}~E, {Furholz}~A, {Laule}~O, {Bleurer}~S,
  {Henning}~L, {Prelic}~A, {Von Rohr}~P, {Thiele}~L, {Zitzler}~E, {Gruissem}~W,
  {Buhlmann}~P. Sparse graphical gaussian modeling of the isoprenoid gene
  network in arabidopsis thaliana. Genome Biology. 2004;\hspace{0pt}5(11):R92+.

\bibitem{Cornillonenglish12}
{Cornillon}~PA, {Guyader}~A, {Husson}~F, {J\'egou}~N, {Josse}~J, {Kloareg}~M,
  {Matzner-L{\o}ber}~E, {Rouvi\`ere}~L. R for statistics. Rennes: Chapman \&
  Hall/CRC Computer Science \& Data Analysis; 2012.

\bibitem{Seaman12}
{Seaman}~SR, {Bartlett}~JW, {White}~IR. {Multiple imputation of missing
  covariates with non-linear effects and interactions: an evaluation of
  statistical methods.} BMC medical research methodology.
  2012;\hspace{0pt}12(1):46.

\bibitem{Bartlett13}
{Bartlett}~JW, {Seaman}~SR, {White}~IR, {Carpenter}~JR. {Multiple imputation of
  covariates by fully conditional specification: accommodating the substantive
  model}. ArXiv e-prints. 2013;\hspace{0pt}In revision.

\bibitem{Green06}
Greenacre~M, Blasius~J. Multiple correspondence analysis and related methods.
  Chapman \& Hall/CRC; 2006.

\bibitem{Kiers91}
{Kiers}~HAL. Simple structure in component analysis techniques for mixtures of
  qualitative and quantitative variables. Psychometrika.
  1991;\hspace{0pt}56:197--212.

\bibitem{pages14}
{Pag\`es}~J. Multiple factor analysis by example using r. Chapman \& Hall/CRC
  The R Series; Taylor \& Francis; 2014;
  \urlprefix\url{http://books.google.fr/books?id=EOxZngEACAAJ}.

\bibitem{Audigier13}
{Audigier}~V, {Husson}~F, {Josse}~J. {A principal components method to impute
  missing values for mixed data}. ArXiv e-prints. 2013;\hspace{0pt}In revision.

\end{thebibliography}
\newpage
\appendix
\section{Simulation study with a block diagonal structure for the covariance matrix - Results for the regression coefficient}
 \begin{table}[h!]
 \caption{Results for the regression coefficient. Root mean squared error for the parameter $\psi=\beta_{X_2}$ estimated by Listwise deletion, Amelia, BayesMI and BayesMIPCA on different configurations varying the number $\Nbind$ of individuals, the number $\Nbvar$ of variables, the correlation $\rho$ between variables and the percentage of missing values. The median confidence interval width for the full data are also provided. For each configuration, 1000 incomplete data sets are generated. Note that $\beta_{X_2}$ can not be estimated if $\Nbind<\Nbvar$. Some values are not available because of failures of the algorithms}
\begin{center}
$\begin{array}{c|c|c|c|c||c|c|c|c|}
\cline{2-9}
&\multicolumn{4}{c||}{\text{parameters}}&\multicolumn{4}{c|}{\text{root mean square error}}\\ \cline{2-9}
&\Nbind&\Nbvar&\rho&\text{\%}&\text{LD}&\text{Amelia}&\text{BayesMI}&\text{BayesMIPCA}\\
\cline{2-9}
1&30&6&0.3&0.1&0.352&0.269&0.249&0.194\\
2&30&6&0.3&0.3&&&0.391&0.183\\
3&30&6&0.9&0.1&0.335&0.277&0.242&0.171\\
4&30&6&0.9&0.3&&&0.362&0.127\\
9&200&6&0.3&0.1&0.099&0.078&0.078&0.066\\
10&200&6&0.3&0.3&0.266&0.115&0.109&0.062\\
11&200&6&0.9&0.1&0.093&0.075&0.074&0.058\\
12&200&6&0.9&0.3&0.265&0.118&0.11&0.046\\
13&200&60&0.3&0.1&&&0.113&0.072\\
14&200&60&0.3&0.3&&&0.171&0.054\\
15&200&60&0.9&0.1&&&0.113&0.072\\
16&200&60&0.9&0.3&&&0.11&0.053\\
\cline{2-9}
\end{array}$
\end{center}
\vspace{1cm}
\caption{Results for the regression coefficient. 95\% coverage and  median confidence interval width for the parameter $\psi=\beta_{X_2}$ estimated by Listwise deletion, Amelia, BayesMI and BayesMIPCA on different configurations varying the number $\Nbind$ of individuals, the number $\Nbvar$ of variables, the correlation $\rho$ between variables and the percentage of missing values. The median confidence interval width for the full data are also provided. For each configuration, 1000 incomplete data sets are generated. Note that $\beta_{X_2}$ can not be estimated if $\Nbind<\Nbvar$. Some values are not available because of fails of the algorithms}
\begin{center}
$\begin{array}{c|c|c|c|c||c|c|c|c|c||c|c|c|c|} \cline{2-14}
&\multicolumn{4}{c||}{\text{parameters}}&\multicolumn{5}{c||}{\text{confidence interval width}}&\multicolumn{4}{c|}{\text{coverage}}\\ \cline{2-14}
&\Nbind&\Nbvar&\rho&\text{\%}&\rotatebox{80}{\text{LD}}&\rotatebox{80}{\text{Amelia}}&\rotatebox{80}{\text{BayesMI}}&\rotatebox{80}{\text{BayesMIPCA}}&\rotatebox{80}{\text{Full data}}&\rotatebox{80}{\text{LD}}&\rotatebox{80}{\text{Amelia}}&\rotatebox{80}{\text{BayesMI}}&\rotatebox{80}{\text{BayesMIPCA}}\\
\cline{2-14}
1&30&6&0.3&0.1&1.332&1.058&0.989&0.936&0.818&0.945&0.94&0.953&0.974\\
2&30&6&0.3&0.3&&&2.492&1.147&0.818&&&0.981&0.997\\
3&30&6&0.9&0.1&1.286&1.051&0.991&0.915&0.791&0.952&0.951&0.957&0.994\\
4&30&6&0.9&0.3&&&2.972&1.108&0.791&&&0.992&1\\
9&200&6&0.3&0.1&0.389&0.313&0.313&0.307&0.278&0.954&0.955&0.96&0.98\\
10&200&6&0.3&0.3&1.011&0.444&0.432&0.359&0.278&0.953&0.945&0.94&0.995\\
11&200&6&0.9&0.1&0.374&0.307&0.306&0.3&0.267&0.956&0.958&0.971&0.99\\
12&200&6&0.9&0.3&0.966&0.465&0.442&0.349&0.267&0.956&0.944&0.949&0.999\\
13&200&60&0.3&0.1&&&0.467&0.373&0.332&&&0.955&0.989\\
14&200&60&0.3&0.3&&&2.716&0.428&0.332&&&1&1\\
15&200&60&0.9&0.1&&&0.465&0.373&0.332&&&0.956&0.993\\
16&200&60&0.9&0.3&&&1.012&0.431&0.332&&&1&1\\
\cline{2-14}
\end{array}$
\end{center}
\end{table}

\end{document}